\documentclass[a4paper,11pt]{article}
\usepackage{jheppub} 
\usepackage[T1]{fontenc} 
\usepackage[utf8]{inputenc}

\usepackage{mathrsfs}  

\usepackage{subcaption} 
\usepackage{cleveref} 
\AddToHook{cmd/appendix/before}{\crefalias{section}{appendix}} 
\usepackage{slashed} 
\usepackage{physics}
\usepackage{placeins} 
\usepackage[normalem]{ulem}
\usepackage{eucal}
\usepackage{hhline}
\usepackage{yfonts}
\usepackage{wasysym}
\def\nn{\nonumber}
\newcommand{\mDm}{m_\chi}

\newcommand{\ana}{\mathcal{A}}

\newcommand{\Lag}{\mathcal{L}} 
\newcommand{\hc}{\text{h.c.}} 

\newcommand{\fm}{\,\text{fm}}

\newcommand{\kev}{\,\text{keV}} 
\newcommand{\mev}{\,\text{MeV}} 
\newcommand{\gev}{\,\text{GeV}} 
\newcommand{\tev}{\,\text{TeV}} 

\newcommand{\arctanh}{\text{arctanh}}

\DeclareMathAlphabet{\mathcal}{OMS}{cmsy}{m}{n}

\graphicspath{{./figures/}}
\title{\boldmath Celestial probes of dark matter electromagnetic interactions}
\author[a]{Alejandro Ibarra}
\author[b]{Merlin Reichard}
\author[c]{Gaurav Tomar}
\affiliation[a]{Technical University of Munich, TUM School of Natural Sciences, Physics Department, 85748 Garching, Germany}
\affiliation[b]{School of Physics, Korea Institute for Advanced Study, Seoul 02455, Republic of Korea}
\affiliation[c]{Center for Quantum Spacetime, Sogang University, 35 Baekbeom-ro, Mapo-gu, Seoul, 121-742, South Korea}
\emailAdd{ibarra@tum.de, mreichard@kias.re.kr, tomar@sogang.ac.kr
}
\preprint{CQUeST-2026-0789}
\abstract{We investigate the neutrino flux from the Sun and Earth, as well as the heating of Jupiter and white dwarfs, induced by the annihilation of spin-$1/2$ dark matter particles captured through their electromagnetic interactions with nuclei. Using current data, we derive constraints on the electromagnetic moments and compare them with existing bounds from direct dark matter searches, laboratory experiments, and astrophysical observations. We find that celestial bodies provide leading constraints in several regions of parameter space: the Earth is particularly sensitive to millicharged dark matter, while massive white dwarfs provide powerful probes of the magnetic dipole moment, charge radius, and anapole moment in regions where direct searches lose sensitivity. We illustrate the implications of these constraints in simplified models of Dirac and Majorana dark matter with radiatively generated electromagnetic moments, as we discuss the complementarity between capture in celestial bodies and other dark matter searches.
}
\begin{document} 
\maketitle
\flushbottom

\section{Introduction}
Numerous observations point towards the existence of an additional matter component in our Universe that interacts extremely weakly with the photon field: the so-called dark matter (DM). Among the many candidates proposed to account for the dark matter of our Universe (see, {\it e.g.}, Refs.~\cite{Jungman:1995df,Bergstrom:2000pn,Bertone:2004pz,Feng:2010gw,Cirelli:2024ssz} for reviews), the weakly interacting massive particle stands out for its simplicity and plausibility. In the simplest scenario, one postulates the existence of a dark matter field that interacts in pairs with Standard Model fields, with a coupling strength sufficiently large to maintain a population of dark matter particles in thermal equilibrium with the Standard Model plasma, yet sufficiently small to allow them to freeze out at early times. The resulting relic abundance can then constitute a significant fraction, or even the entirety, of the dark matter abundance observed today.

If the dark matter field is a fundamental spin-1/2 fermion, four-point interactions involving two dark matter fields and two Standard Model fields necessarily induce at the quantum level interactions between the dark matter and the photon field~\cite{Flacke:2006ut,Kopp:2014tsa,Ibarra:2015fqa,Sandick:2016zut,Baker:2018uox,Hisano:2018bpz,Herrero-Garcia:2018koq,Kavanagh:2018xeh,Hisano:2020qkq,Ibarra:2022nzm,Arcadi:2023imv,Ibarra:2024mpq}. Moreover, if the dark matter particle is a composite state, electromagnetic interactions of its constituents can likewise induce a coupling to the photon, in close analogy with the neutron magnetic moment, which ultimately originates from the electric charges of its constituent quarks~\cite{Bagnasco:1993st,LatticeStrongDynamicsLSD:2013elk,Antipin:2015xia,Aranda:2015jis,Asadi:2026mip}.

From a model-independent perspective, the electromagnetic interactions of a spin-$1/2$ fermion at low momentum transfer can be parameterized in terms of five electromagnetic properties: the electric charge, charge radius, magnetic and electric dipole moments, and anapole moment. A coupling of dark matter to the photon through any of these electromagnetic interactions can induce scattering off nuclei in dedicated direct-detection experiments, potentially allowing dark matter to be identified as a particle and providing information about its underlying properties. The implications of dark matter electromagnetic form factors for direct-detection experiments have been extensively investigated in, {\it e.g.}, Refs.~\cite{Pospelov:2000bq,Sigurdson:2004zp,Masso:2009mu,Fitzpatrick:2010br,Banks:2010eh,Barger:2010gv,Fortin:2011hv,DelNobile:2014eta,Gresham:2014vja,Hambye:2021xvd,Ibarra:2022nzm,Ibarra:2024mpq,Kumar:2025wsp}

In addition to these Earth-based experiments, celestial bodies have been proposed as complementary probes of particle dark matter~\cite{Gould:1987ir,Press:1985ug,Mack:2007xj,Feng:2015hja,Feng:2016ijc,Chauhan:2016joa,Brenner:2020mbp,Curtin:2020tkm,Panotopoulos:2020kuo,Leane:2020wob,Bell:2021fye,Bose:2021yhz,Leane:2021tjj,Bramante:2022pmn,Li:2022wix,Leane:2024bvh,Bramante:2023djs,Bose:2023yll,Croon:2023bmu}. In this work, we investigate signatures arising from the capture of spin-1/2 dark matter particles through their electromagnetic interactions and their subsequent annihilation into Standard Model particles. Concretely, we study the neutrino flux produced by dark matter annihilation in the centers of the Sun and the Earth, as well as the heating induced by dark matter annihilation in Jupiter and in massive white dwarfs in the M4 cluster (see, \textit{e.g.}, Refs.~\cite{Curtin:2020tkm,Panotopoulos:2020kuo,Bell:2021fye} for related studies).
Owing to the non-trivial dependence of the scattering rate on the different electromagnetic form factors, we expect a strong complementarity between direct-detection experiments, which predominantly probe the high-velocity tail of the dark matter velocity distribution, and capture in celestial bodies, which is particularly sensitive to its low-velocity tail.

This work is structured as follows: in \cref{sec:capture_rate} we calculate the scattering rate of dark matter particles off nuclei mediated by electromagnetic multipoles, as well as the capture rate in a celestial body. In \cref{sec:MI-constraints} we determine the limits on the various electromagnetic moments from the non-observation of a neutrino excess in the direction of the center of the Sun or the Earth, and from the requirement that the heat emission of Jupiter or a heavy white dwarf does not exceed their measured values. In Sections
\ref{sec:Dirac-toy-model} and 
\ref{sec:Majorana-toy-model} we discuss the implications of our constraints on the parameter space of a toy model with Dirac or Majorana dark matter with a scalar mediator. Finally in  \cref{sec:Conclusion} we present our conclusions. We also include Appendix \ref{app:loop_functions}, containing analytical formulas of the electromagnetic moments of a neutral spin-1/2 fermion that are generated 
at the one-loop level by the interactions with a scalar mediator and a Standard Model fermion.

\section{Dark matter Capture through electromagnetic multipoles}\label{sec:capture_rate}

The low-energy interaction between a spin-1/2 DM candidate $\chi$ with the photon can be described by the Lagrangian (see \textit{e.g.} Ref.~\cite{Broggini:2012df})\
\begin{align}\label{eq:EM_Lagrangian}
\Lag_{\rm int}=
e Q_\chi  \bar\chi \gamma^\mu \chi A_\mu +
\frac{\mu_\chi}{2} \bar \chi \sigma^{\mu\nu} \chi F_{\mu\nu} +
\frac{d_\chi}{2} i \bar \chi \sigma^{\mu\nu} \gamma^5 \chi F_{\mu\nu}+ b_\chi \bar \chi \gamma^\mu \chi \partial^\nu F_{\mu\nu} +
\mathcal{A}_\chi \bar \chi \gamma^\mu \gamma^5 \chi \partial^\nu F_{\mu\nu},
\end{align}
where $A^\mu$ is the standard model (SM) photon field, and $F^{\mu\nu}$ the $\text{U}(1)_\text{EM}$ field strength tensor. The coefficients in \cref{eq:EM_Lagrangian} are the millicharge $Q_\chi$,  magnetic dipole moment (MDM) $\mu_\chi$, electric dipole moment (EDM) $d_\chi$, charge radius $b_\chi$, and anapole moment $\ana_\chi$.

The dark matter electromagnetic interactions lead to the scattering with nuclei. For a target nucleus with mass $m_T$ and spin $J_T$ the differential scattering cross section reads:
\begin{align}
\frac{d\sigma_T}{dE_R}
={}&
\frac{2m_T}{(2J_T+1)v_T^2}
\sum_{\tau,\tau'=n,p}
\Bigg\{
\left[c_1^\tau c_1^{\tau'}+\frac{1}{4}v_T^{\perp 2}\left(\frac{q^2}{m_N^2}c_5^\tau c_5^{\tau'}+c_8^\tau c_8^{\tau'}
\right)+\frac{q^2}{4m_N^2}c_{11}^\tau c_{11}^{\tau'}\right]W_{TM}^{\tau\tau'}
\nonumber\\
&+
\frac{1}{16}\left[c_4^\tau c_4^{\tau'}
+\frac{q^2}{m_N^2}c_9^\tau c_9^{\tau'}\right]W_{T\Sigma'}^{\tau\tau'}+\frac{1}{16}\left[c_4^\tau+\frac{q^2}{m_N^2}c_6^\tau\right]\left[c_4^{\tau'}+\frac{q^2}{m_N^2}c_6^{\tau'}\right]
W_{T\Sigma''}^{\tau\tau'} \nonumber
\\
&+\frac{q^2}{4m_N^2}
\left[
\frac{q^2}{m_N^2}c_5^\tau c_5^{\tau'}+c_8^\tau c_8^{\tau'}\right]
W_{T\Delta}^{\tau\tau'}
+
\frac{q^2}{4m_N^2}\left[c_5^\tau c_4^{\tau'}-c_8^\tau c_9^{\tau'}\right]
W_{T\Delta\Sigma'}^{\tau\tau'}
\Bigg\},
\end{align}
where $q^2=2m_T E_R$ is the squared momentum transfer and $v_T^{\perp 2}=v_T^2-q^2/(4\mu_{\chi T}^2)$ is the squared transverse velocity, with $v_T$ denoting the relative velocity between the dark matter particle and the target nucleus. Furthermore, the parameters $c_i$ denote the Wilson coefficients of the operators $\mathcal{O}_i$ of the non-relativistic effective field theory (NREFT) for dark matter--nucleon interactions constructed in Ref.~\cite{Fitzpatrick:2012ix}. A subset of these operators arises when the photon in the Lagrangian of \cref{eq:EM_Lagrangian} is coupled to the electromagnetic nucleon current. Concretely, the following Wilson coefficients are generated~\cite{DelNobile:2018dfg}:
\begin{equation}\label{eq:NR_coefficients}
	\begin{gathered}
			c_1^{N}=  \frac{e^2 Q_N}{q^2} Q_\chi +  \frac{2e Q_N }{4 m_\chi}\mu_\chi +e Q_N  b_\chi\;,\\
			c_4^N= \frac{ 2e g_N }{2 m_N}\mu_\chi  \;,  \quad
			c_5^{N}=  \frac{ 2e Q_N m_N}{q^2} \mu_\chi \;, \quad
			c_6^N= -\frac{2e g_Nm_N}{2 q^2} \mu_\chi  \;,\\
			c_8^{N}=  2e Q_N \mathcal{A}_\chi \;,  \quad
			c_9^{N}= -e g_N \mathcal{A}_\chi  \;, \quad
			c_{11}^N=\frac{2e Q_N m_N }{q^2}  d_\chi \;,
		\end{gathered}    
	\end{equation}
where  $Q_p = 1$ and $Q_n = 0$ are the proton and neutron charge, respectively, whereas $g_p = 5.59$ and $g_n = -3.83$ are their $g$-factors.  Finally, $W_{TM}$, $W_{T\Sigma'}$, $W_{T\Sigma''}$, $W_{T\Delta}$, and $W_{T\Delta\Sigma'}$ are nuclear response functions for the target $T$, which can be found in Ref.~\cite{Fitzpatrick:2012ix}. Roughly, $W_{TM}$ describes the proton and neutron distributions within the target nucleus, $W_{T\Sigma'}$ and $W_{T\Sigma''}$ encode the nuclear spin components transverse and longitudinal to the momentum transfer, respectively, and $W_{T\Delta}$ describes the response associated with the orbital angular momentum of the nucleons. Finally, $W_{T\Delta\Sigma'}$ accounts for the interference between the orbital and transverse-spin responses.

It is important to note that the contribution of the electric charge to the differential cross section is inversely proportional to $E_R$, suggesting that the total cross section diverges as $E_R\rightarrow 0$. In a celestial body, however, the nucleus is not isolated, but is surrounded by electrons. At sufficiently small momentum transfer, corresponding to small nuclear recoil energies, the dark matter particle cannot resolve the nucleus independently of its electron cloud and instead probes the atom as a whole. Since the atom is electrically neutral at sufficiently large distances, the nuclear charge is screened by the electron cloud. Consequently, the effective charge probed by the dark matter particle decreases as $q\rightarrow 0$, regularizing the apparent infrared divergence and yielding a finite cross section in this limit. The scattering at very small momentum transfer can then be dominated by the nuclear magnetic response rather than by the screened electric charge response.

To account for the screening of the nuclear charge by the surrounding electrons, we modify the proton charge response using the Thomas-Fermi approximation
\begin{align}
W_{TM}^{pp}(q)\;\longrightarrow\;W_{TM}^{pp}(q)\left(\frac{q^2}{q^2+k_{\rm TF}^2}\right)^2 ,
\label{eq:screening}
\end{align}
with $k_{\rm TF}$  the screening momentum, given by~\cite{jancovici1962relativistic}
\begin{equation}
k_{\rm TF}^2=\frac{4\alpha}{\pi}k_F E_F =\frac{4\alpha}{\pi}k_F\sqrt{k_F^2+m_e^2},
\label{eq:kTF_rel}
\end{equation}
where $k_F=(3\pi^2n_e)^{1/3}$ is the electron Fermi momentum and $E_F=\sqrt{k_F^2+m_e^2}$ its Fermi energy. This expression applies also when the electrons are relativistic, as is the case in the interior of a white dwarf. For our benchmark electron density, $n_e\simeq3\times10^{-6}\,\fm^{-3}$~\cite{Bell:2021fye}, one finds $k_F\gg m_e$, and \cref{eq:kTF_rel} approaches the ultra-relativistic result $k_{\rm TF}^2\simeq(4\alpha/\pi)k_F^2$.

In the non-relativistic limit, $k_F\ll m_e$, \cref{eq:kTF_rel} reduces to
\begin{equation}
k_{\rm TF}^2\simeq\frac{4\alpha m_e k_F}{\pi}.
\label{eq:kTF_NR}
\end{equation}
For electrons bound in an atom, estimating their characteristic Fermi momentum in the Thomas-Fermi approximation leads to the conventional atomic screening length
\begin{equation}
a=k_{\rm TF}^{-1}=\frac{1}{4}\left(\frac{9\pi^2}{2Z}\right)^{1/3}a_0,
\end{equation}
with $a_0$ the Bohr radius. We therefore employ the relativistic expression in \cref{eq:kTF_rel} for white dwarfs, while for ordinary non-relativistic atomic matter we use the corresponding Thomas-Fermi atomic screening length.

At a given position in our Galaxy, characterized by a dark matter density $\rho_\chi$ and velocity distribution $f(\vec v)$, dark matter particles can scatter off nuclei in celestial bodies. The differential scattering rate is given by
\begin{align}
\frac{d R_{\chi T}}{d E_R}=\sum_T N_T \int_{v_{\rm min}} d^3 v_T\,\frac{\rho_{\chi}}{m_{\chi}}\,f(\vec{v}_T)\,v_T\,\frac{d\sigma_T}{d E_R},
\label{eq:dr_der_ER_vT}
\end{align}
where $N_T$ denotes the number of nuclear targets of species $T$, and
\begin{equation}
v_{\rm min}=\sqrt{\frac{m_T E_R}{2\mu_{\chi T}^2}},
\end{equation}
with $\mu_{\chi T}$ the reduced mass of the dark matter particle and the  target nucleus $T$.

The DM capture rate in a spherically symmetric celestial body of radius $R$ is given by~\cite{Gould:1987ir},
\begin{equation}\label{eq:capture_rate_general}
    C = \frac{\rho_\chi}{\mDm} \int_{0}^{R} dr\,r^2\, 4\pi \sum_T \frac{dC_T(r)}{dV}, 
\end{equation}
where the sum $\sum_T$ runs over the targets $T$ in the celestial body. The differential capture rate reads, 
\begin{equation}\label{eq:capture_rate_general_diff}
    \frac{dC_T(r)}{dV} = \int du\, f(u) \frac{1}{u} w(u,r)^2\, \eta_T(r) \,\Theta(E_\text{max}^\chi -E_\text{cap}^\chi) \int_{E^\chi_\text{cap}(u)}^{E_\text{max}^\chi(u,r)} d E_R\, \frac{d\sigma_T}{dE_R},
\end{equation}
where $\eta_T(r)$ is the number density of target nuclei in the celestial body, encoding its nuclear composition. The first integral is performed over the asymptotic speed $u$ of the incoming DM particle far from the center of the celestial body. The DM speed at a radial position $r$ is related to $u$ through,
\begin{equation}
    w^2(u,r)=u^2+v_{\rm esc}^2(r),
\end{equation}
where $v_{\rm esc}(r)$ is the local escape velocity. The recoil-energy integral extends from the minimum energy required for capture,
\begin{equation}
    E_{\rm cap}^{\chi}(u)=\frac{1}{2}\mDm u^2,
\end{equation}
to the maximum kinematically allowed recoil energy,
\begin{equation}
    E_{\rm max}^{\chi}(u,r) =\frac{1}{2}\mDm w^2 \left[ 1-\frac{\mu_{\chi T}^2}{m_T^2}\left(1-\frac{m_T}{\mDm}\right)^2 \right],
\end{equation}
with  $\mu_{\chi T}=m_\chi m_T/(m_\chi+m_T)$ is the reduced mass of the DM particle and the target nucleus.

Captured dark matter particles accumulate in the interior of the celestial body, leading to an increase in their number density. As the number density  increases, so does the dark matter annihilation rate, thereby reducing the number of particles accumulated in the celestial body. For dark matter masses for which evaporation can be neglected, the time evolution of the number of captured dark matter particles is governed by
\begin{equation}
    \frac{dN_\chi}{dt}=C-C_A N_\chi^2,
    \label{eq:dN/dt}
\end{equation}
where $C$ is the capture rate and $C_A$ is the annihilation coefficient, given by
\begin{equation}
    C_A = \frac{\langle \sigma v\rangle}{V_{\rm eff}}.
\end{equation}
where $V_{\rm eff}$ is~\cite{Griest:1986yu}:
\begin{equation}
    V_{\rm eff}=
    \frac{\left[4\pi\int_0^{R} r^2 n_\chi(r)\,dr\right]^2}
    {4\pi\int_0^{R} r^2 n_\chi^2(r)\,dr},
\end{equation}
Here $n_\chi(r)$ is the thermalized DM density profile,
\begin{align}
    n_\chi(r) &= \frac{1}{\pi^{3/2}r_\chi^3}
    \exp\left(-\frac{r^2}{r_\chi^2}\right),\\
    r_\chi^2 &= \frac{3k_B T_c}{2\pi m_\chi G \rho_c},
\end{align}
with $k_B$ is the Boltzmann constant, $G$ is the gravitational constant, and $\rho_c$ and $T_c$  the core mass density and central temperature of the celestial body, respectively.

The number of captured dark matter particles as a function of time can be straightforwardly calculated from \cref{eq:dN/dt}, the result being
\begin{equation}
N_\chi(t)=\sqrt{\frac{C}{C_A}}\,
\tanh\left(\frac{t}{\tau_{\rm eq}}\right)
\end{equation}
where $ \tau_{\rm eq}=
1/\sqrt{C C_A}$ is the equilibration time between capture and annihilation and hence the total dark matter annihilation rate is
\begin{equation}
\Gamma_A(t)=\frac{1}{2}C_A N_\chi^2(t)=\frac{C}{2}\tanh^2\left(\frac{t}{\tau_{\rm eq}}\right).
\end{equation}

In our analysis, we  utilize the public code \texttt{WimPyDD}~\cite{Jeong:2021bpl} and its extension \texttt{WimPyC}~\cite{Kang:2025yci}
for the calculation of the capture- and annihilation rate.

\section{Probing dark matter electromagnetic multipoles from annihilations}
\label{sec:MI-constraints}
The annihilation of dark matter particles captured in celestial bodies can give rise to observable signatures. In our analysis, we consider the neutrino flux produced by dark matter annihilations in the Sun and the Earth, as well as the heating induced by dark matter annihilations in Jupiter and in white dwarfs. For concreteness, we consider annihilations into $\tau^+\tau^-$ and $W^+W^-$ final states, which yield relatively hard neutrino spectra, and adopt an annihilation cross section $\langle \sigma v\rangle=3\times10^{-26}\,{\rm cm^3\,s^{-1}}$. 

For the calculation of the capture and annihilation rates, and for the comparison with observational data, we adopt the following assumptions for the different celestial bodies:
\begin{itemize}
    \item \textbf{Sun:} We assume the standard solar model 
    AGSS09ph~\cite{Serenelli:2009yc} for the composition of the Sun, and a velocity distribution given by:
\begin{align}
f_\odot(\mathbf v)=\frac{1}{N(\pi v_0^2)^{3/2}}\exp\left[-\frac{|\mathbf v+\mathbf v_\odot|^2}{v_0^2}\right]
\label{eq:vel_distribution}
\end{align}
where $N$ is a normalization factor. This corresponds to a Maxwell-Boltzmann distribution truncated at the Galactic escape velocity, $v_{\rm esc}=550\,\text{km/s}$, expressed in the rest frame of the Sun, which moves through the Galactic halo with velocity $v_\odot=220\,\text{km/s}$. 
Finally,   We derive constraints on the electromagnetic  multipole moments using the upper limits on the neutrino flux from the Sun obtained by IceCube~\cite{IceCube:2016dgk} and DeepCore~\cite{IceCube:2021xzo}. Concretely, we use the limits shown in fig.~5 of Ref.~\cite{Choi:2024rmq} for IceCube and those reported in Table~IV of Ref.~\cite{IceCube:2021xzo}  for DeepCore. 
\item \textbf{Earth:} We assume the Preliminary Reference Earth Model of Ref.~\cite{DZIEWONSKI1981297}, with the chemical   composition taken from Ref.~\cite{Bramante:2019fhi}, and a velocity distribution given by \cref{eq:vel_distribution}. We then derive constraints on the electromagnetic multipole moments from the upper limits on the neutrino flux inferred from the $90\%$ C.L. limits on the standard spin-independent dark matter--nucleon cross section reported by IceCube using its 10-year data set~\cite{IceCube:2024yaw}.
\item \textbf{Jupiter:} We assume the Jovian
model J11-4a for the density profile~\cite{french2012ab} with a composition of 75$\%$ hydrogen and 25$\%$ helium. For the range of  parameters considered in this work, however, the capture rate is approximately  saturated at its geometric value. We then compare the luminosity generated by dark matter annihilations with Jupiter's measured internal heat flux, determined from Cassini data to be $L_{\jupiter}/A_{\jupiter}= (7.485\pm0.163)\,{\rm W\,m^{-2}}$, where $A_{\jupiter}$ denotes Jupiter's surface area and $L_{\jupiter}$ its internal luminosity~\cite{li2018less}. We assume that all the energy released in dark matter annihilations is deposited in the planet.
\item \textbf{White dwarfs:} We closely follow   Ref.~\cite{Bell:2021fye} for the capture, considering a white dwarf with a mass of $1.384$ M$_{\odot}$ and radius of  $1.25\times 10^3$ km. The density profile is taken assuming a 100$\%$ carbon composition. To set constraints on the multipole moments, we require that the luminosity generated by dark  matter annihilations in the massive white dwarfs of the M4 globular cluster  does not exceed the observed luminosity, $L^{\rm WD}=2.95\times10^{31}\,{\rm GeV\,s^{-1}}$. The dark matter density in M4 is estimated assuming a compressed NFW profile, with a local density at the position of the white dwarf of  $\rho_\chi=798\,{\rm GeV\,cm^{-3}}$  following Ref.~\cite{McCullough:2010ai}. The dark matter particles are assumed  to follow a Maxwell-Boltzmann velocity distribution characterized by $v_{\rm esc}=1000\,{\rm km\,s^{-1}}$ and $v_{\rm rms}=8\,{\rm km\,s^{-1}}$ following Ref.~\cite{Bell:2021fye}.
\end{itemize}

\begin{figure}
    \centering
    \includegraphics[width=0.49\linewidth]{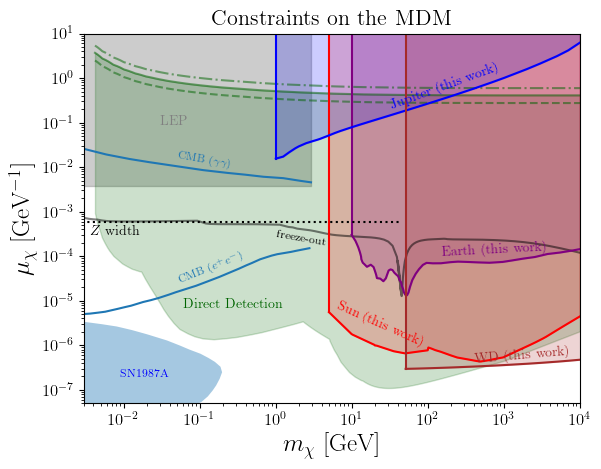}
    \includegraphics[width=0.49\linewidth]{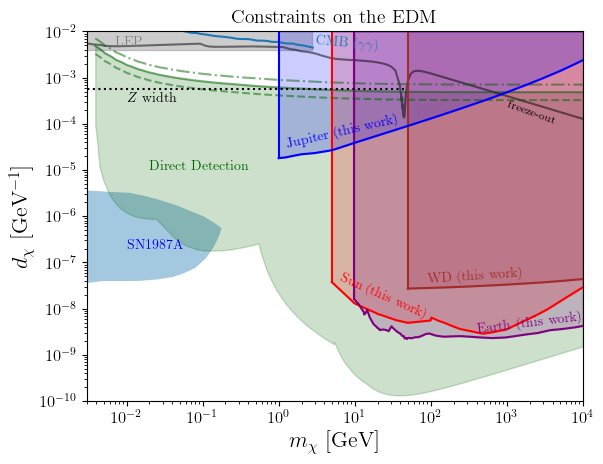}\\
    \includegraphics[width=0.49\linewidth]{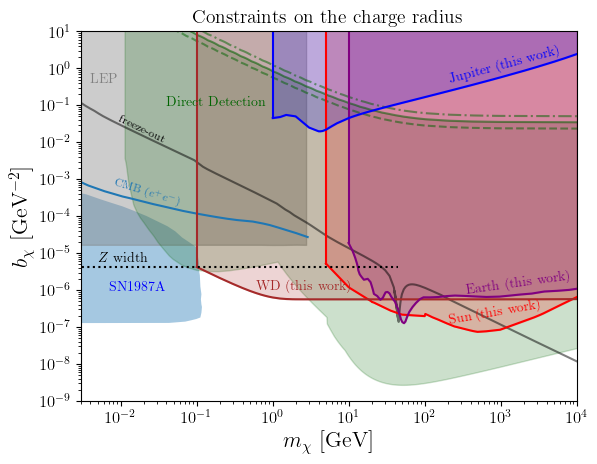}
    \includegraphics[width=0.49\linewidth]{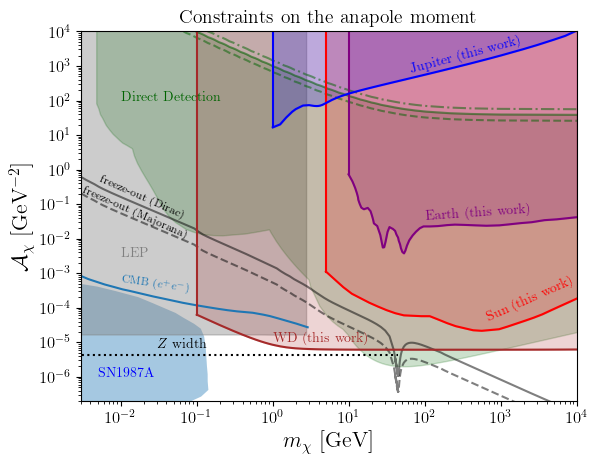}\\
    \includegraphics[width=0.49\linewidth]{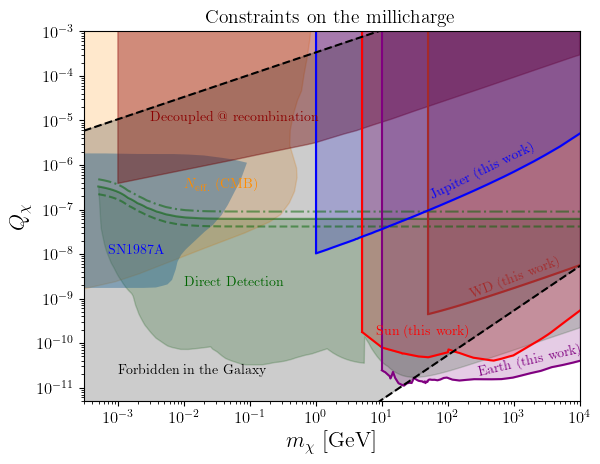}
    \caption{Constraints on the individual electromagnetic moments of a spin-1/2 Dirac DM candidate inferred from celestial bodies as described in the main text. We include the dominant limit from direct detection experiments, taken from Ref.~\cite{Ibarra:2024mpq}, with estimated attenuation ceilings corresponding to  $\ell = \{0.5, 1, 2 \}$ km of overburden in Earth's crust.}
    \label{fig:exclusion}
\end{figure}

We show in \cref{fig:exclusion} the regions in the exclusion limits for the various electromagnetic moments, assuming that only one of them is  non-zero at a time. For each celestial body, we only consider mass values that are above the threshold at which evaporation becomes relevant. Concretely,  for the Sun we set $m_\text{evap.}\sim 4\gev$~\cite{Garani:2021feo}, for Jupiter  $m_\text{evap.} \sim 1\gev$~\cite{Li:2022wix,French:2022ccb}, for Earth, $m_\text{evap.} \sim 10\gev$ ~\cite{Garani:2021feo}, and for heavy white dwarfs $m_\text{evap.}\sim 65\kev$~\cite{Bell:2021fye}. For the heavy white dwarfs we consider as lower bound $\mDm \gtrsim 100\mev$ ($50\gev$) for the anapole and charge radius (millicharge, MDM, EDM), as collective effects are expected to alter the standard capture process below these values~\cite{DeRocco:2022rze}. The plot also shows the excluded region of the parameter space from direct detection experiments; the lower limit of the excluded region is determined by the non-observation of a dark matter signal at direct detection experiments, while the upper limit, from the attenuation of the dark matter flux by the rock for an overburden on 0.5, 1 and 2 km (from bottom to top), and which prevents dark matter particles from reaching the detector. 

We also show cooling constraints from supernova 1987A~\cite{Chang:2018rso,Chu:2018qrm} (for the millicharge we use the constraints assuming the ``fiducial'' density profile), LEP~\cite{Chu:2018qrm}, CMB~\cite{Chu:2018qrm}, and $N_\text{eff}$~\cite{Vogel:2013raa}. For millicharge, we include the constraints that stem from requiring that DM is decoupled at the time of recombination, as well as the upper limit on the millicharge such that galactic magnetic field prevents millicharged particles from entering the Galaxy, and the lower limit such that the millicharged  that are already present in the Galaxy can be blown away from the disk by supernovae \cite{McDermott:2010pa}. 
The dotted horizontal lines indicate the exclusion limit from the $Z$-width measurement in a model in which DM couples to the hypercharge gauge boson $B_\mu$~\cite{Chu:2018qrm} (see Ref.~\cite{Arina:2020mxo} for a discussion of the high-energy divergent behavior arising in a photon-only coupled theory). For the dimension-full electromagnetic interactions, we include the relic density constraints, obtained by fixing the respective parameter to satisfy $\Omega_\text{DM} h^2 = 0.12$~\cite{Planck:2018vyg}, where we used \texttt{micrOMEGAs}~\cite{Belanger:2013oya} to calculate the model-prediction via thermal freeze-out. The resonance around $\mDm\simeq m_Z/2$ is due to the fact that we assumed a hypercharge-coupled DM candidate that we implemented as \texttt{FeynRules}~\cite{Alloul:2013bka} model file. For the millicharge, the freeze-out line is mostly above the shown parameter ranges and thus fully excluded by cosmological constraints.

We find that the limits derived via the celestial bodies can impose the strongest constraints for some of the electromagnetic moments. In particular, for the dimension-6 operators, the WDs provide the most sensitive probe in the range $10\mev \lesssim \mDm \lesssim 10\gev$; for the anapole operator, they also provide leading constraints for $\mDm \gtrsim 1\tev$. Furthermore, we find that the limits derived using the Earth as target provide leading constraints for DM masses above around $10\gev$ for the millicharge operator. This result is due to the long-range nature of the interaction ($\sim 1/q^2$), which favors capture at Earth and is further enhanced by the presence of heavy elements such as iron, whose large number of protons increases the cross section coherently. For the dimension-5 operators, on the other hand, we do not find an improvement over the direct detection experiments, with the exception of the MDM, where the WD constraints become leading for $\mDm \gtrsim 1\tev$; for the EDM, none of the celestial-body constraints exceed the existing bounds.

We further note that among the celestial bodies, for the MDM, the Sun provides stronger constraints than the Earth, whereas for the EDM, it is the other way around. Similar to the millicharge case, the long-range nature of the EDM interaction ($\sim 1/q$) enhances the capture rate at Earth, and thus improves the constraints over the Sun-based analysis. The dim-6 operators describe contact interactions, explaining the resonance features whenever the DM mass is comparable to one of the dominant elements of the Earth or Jupiter. Meanwhile, for the dipole moments and the millicharge operators, the $q$ dependence obscures this information about the target, resulting in no clear peaks.

\section{Dirac dark matter with a $t$-channel scalar mediator}
\label{sec:Dirac-toy-model}

A common ultraviolet completion of dark matter scenarios 
is the so-called $t$-channel model, in which the DM couples to the SM at tree level through a Yukawa coupling 
with a mediator (see Refs.~\cite{Garny:2013ama,Ibarra:2015fqa,Arcadi:2023imv,Arina:2025zpi,Biondini:2025gpg,Becker:2026icc,Binder:2025daq,Belfatto:2025ids} for related studies). Here we consider the specific scenario where a Dirac dark matter particle couples to the third generation of leptons via a scalar mediator, \textit{i.e.}, a $\tau$-philic dark matter candidate. After diagonalization of the singlet and doublet scalar components, the relevant portal interaction Lagrangian between the dark matter particle $\chi$ (with mass $\mDm$), the SM $\tau$ lepton, and the lightest scalar mediator field $S_1$ (with soft mass $m_{S_1}$) may be written as~\cite{Ibarra:2024mpq}
\begin{equation}\label{eq:L_t_channel}
    \Lag \supset \bar{\chi} \left[c \cos\theta_\text{P} P_L + c\sin\theta_\text{P} e^{i\phi_\text{CP}} P_R\right] S_1^* \tau + \hc,
\end{equation}
with the effective Yukawa coupling $c$, the P-violating angle $\theta_\text{P}$, paremeterizing the coupling strength to the left-and right chirality of $\tau$ as $c_L =c \cos\theta_\text{P}$ and $c_R = c \sin\theta_\text{P}$, and the CP-violating angle $\phi_\text{CP}$. A similar interaction with the heavier scalar $S_2$ is present, however, we will suppose that $S_2$ is sufficiently heavy to contribute negligibly to the electromagnetic moments.

At the one-loop level, the interaction in \cref{eq:L_t_channel} generates all dimension-full electromagnetic moments of $\chi$~\cite{Kopp:2014tsa,Sandick:2016zut,Hisano:2018bpz,Herrero-Garcia:2018koq,Ibarra:2024mpq}. The magnetic dipole moment reads  
\begin{equation}
    \mu_\chi = -\frac{e Q_f }{32\pi^2\mDm} c^2 \left[\mathcal{F}_1\left(\frac{m_\tau}{\mDm},\frac{m_{S_1}}{\mDm}\right)+\sin(2\theta_\text{P})\cos\phi_\text{CP} \mathcal{F}_2\left(\frac{m_\tau}{\mDm},\frac{m_{S_1}}{\mDm}\right)\right],
\end{equation}
the electric dipole moment
\begin{equation}
    d_\chi = \frac{eQ_f }{32\pi^2 \mDm} c^2 \sin(2\theta_\text{P}) \sin\phi_\text{CP} \mathcal{F}_2\left(\frac{m_\tau}{\mDm},\frac{m_{S_1}}{\mDm}\right) ,  
\end{equation}
the charge radius
\begin{equation}
    b_\chi = -\frac{eQ_f}{384\pi^2\mDm^2} c^2 \left[\mathcal{F}_4\left(\frac{m_\tau}{\mDm},\frac{m_{S_1}}{\mDm}\right)+ \cos(2\theta_\text{P}) \cos\phi_\text{CP}\mathcal{F}_5\left(\frac{m_\tau}{\mDm},\frac{m_{S_1}}{\mDm}\right)\right], 
\end{equation}
and the anapole moment
\begin{equation}\label{eq:anapole_moment}
    \mathcal{A}_\chi = -\frac{e Q_f}{192\pi^2\mDm^2} c^2 \cos(2\theta_\text{P}) \mathcal{F}_3\left(\frac{m_\tau}{\mDm},\frac{m_{S_1}}{\mDm}\right). 
\end{equation}
The loop functions $\mathcal{F}_i(\mu,\eta)$ are provided in \cref{app:loop_functions}.

\begin{figure}
    \centering
    \includegraphics[width=0.49\linewidth]{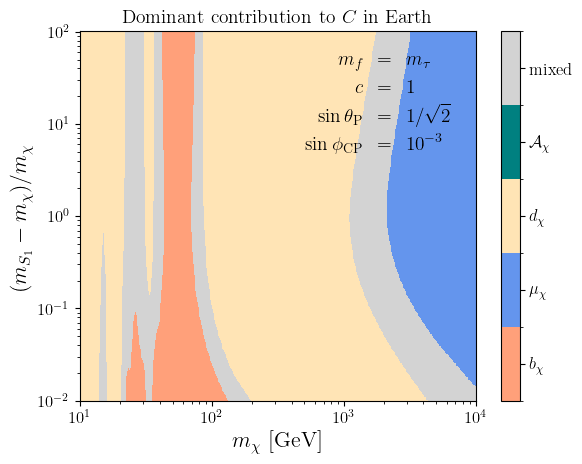}
    \includegraphics[width=0.49\linewidth]{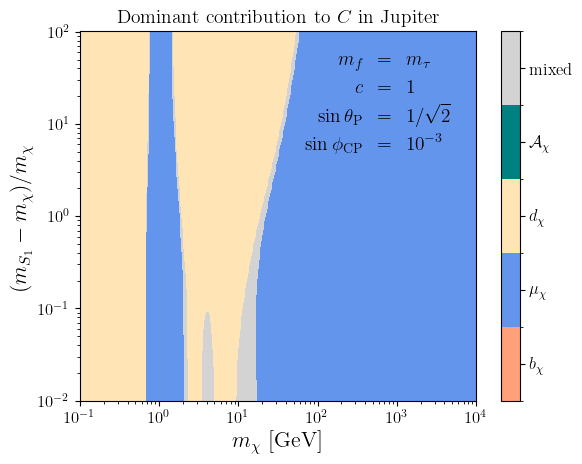}
    \includegraphics[width=0.49\linewidth]{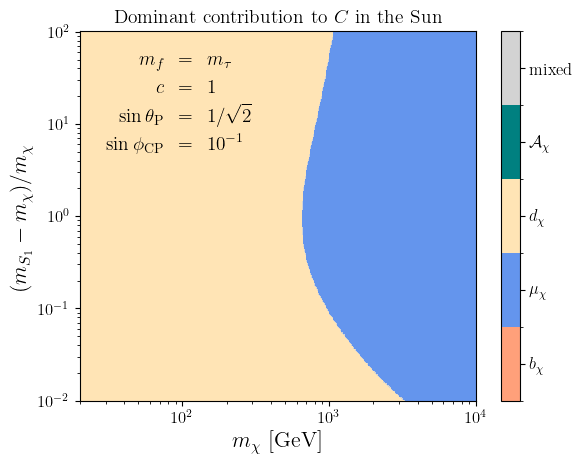}
    \includegraphics[width=0.49\linewidth]{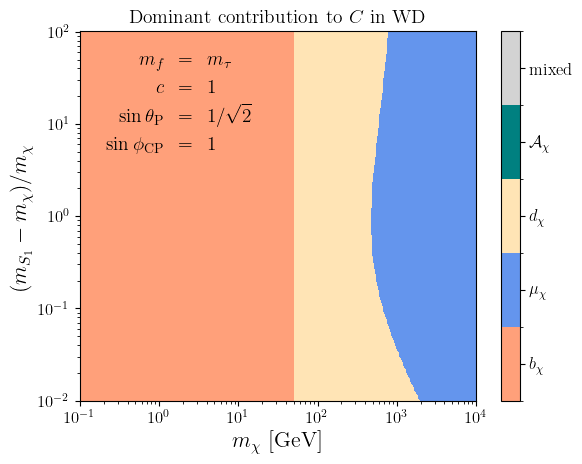}
    \caption{Dominant multipole moment contribution for the  capture in celestial bodies for the $\tau$-philic toy model, assuming maximal P violation, and for different values of the CP violating phase:  capture in Earth for $\sin\phi_{\rm CP}=10^{-3}$ (top left), capture in Jupiter for $\sin\phi_{\rm CP}=10^{-3}$ (top right), capture in the Sun for $\sin\phi_{\rm CP}=10^{-1}$ (bottom left), capture in white dwarfs for $\sin\phi_{\rm CP}=1$ (bottom right).}
    \label{fig:contribution_earth_tau}
\end{figure}

We show in \cref{fig:contribution_earth_tau} the dominant contribution to the capture rate in celestial bodies, for a fixed value of the coupling, $c=1$, assuming maximal parity violation, $\sin\theta_P=1/\sqrt{2}$, and for different values of $\sin\phi_{\rm CP}$, that parametrizes the CP violation in the dark sector. For all celestial bodies, we find a relatively strong sensitivity to the electric dipole moment and thus CP violation. For the Earth and $\sin\phi_\text{CP}=10^{-3}$ (top left panel), we find large regions in the range $20\gev\lesssim\mDm\lesssim100\gev$ in which the charge radius operator dominates the capture at the Earth, even for large mass splittings of $m_{S_1} \simeq 100\,\mDm$. Meanwhile for $\sin\phi_\text{CP}=10^{-2}$, the whole shown region is dominated by the EDM, whereas for a smaller CP angle, $\sin\phi_\text{CP}=10^{-4}$, almost all of the parameter space is dominated by the charge radius operator except for $\mDm \gtrsim 1\tev$, where the MDM drives the capture rate.

For Jupiter (top right panel) with $\sin\phi_\text{CP}=10^{-3}$, we find that the dipole moments contribute the most, with visible resonance peaks when the DM mass is comparable to that of hydrogen or helium. For DM masses above $\sim 20-50 \gev$, the MDM provides the dominant contribution. For larger CP violation of $\sin\phi_\text{CP}=10^{-2}$, the EDM is found to dominate for $\mDm \lesssim 1\tev$, and even $\mDm \lesssim 5\tev$ for small mass-splittings $\eta \sim 1.01$. For $\sin\phi_\text{CP}=10^{-4}$ on the other hand, the whole considered parameter space is dominated by the MDM.

For the capture in the Sun (bottom left), the rate is dominated by the dipole moments. In particular, we find a strong sensitivity to the MDM; for example, a CP-violating angle of $\sin\phi_\text{CP}=10^{-2}$ makes the EDM contribution subdominant. We find a similar behavior for white dwarfs (bottom right), where the amount of CP violation dictates whether the EDM or the MDM dominates the capture rate in the range $50\gev \lesssim \mDm \lesssim 500 - 2000\gev$: for $\sin\phi_\text{CP}=1$, this region is dominated by the EDM, whereas for $\sin\phi_\text{CP}=10^{-1}$, the whole considered parameter space with $\mDm > 50\gev$ is dominated by the MDM. Similar to before, we set $\mu_\chi = d_\chi =0$ for $\mDm < 50\gev$ to remove the region in which long-range effects are estimated to become relevant.

\begin{figure}
    \centering
    \includegraphics[width=0.49\linewidth]{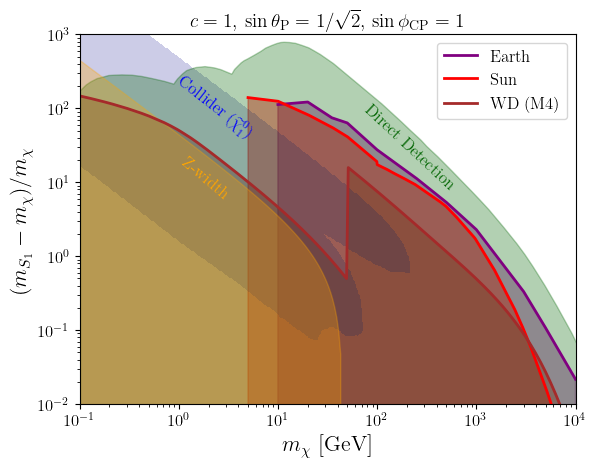}
    \includegraphics[width=0.49\linewidth]{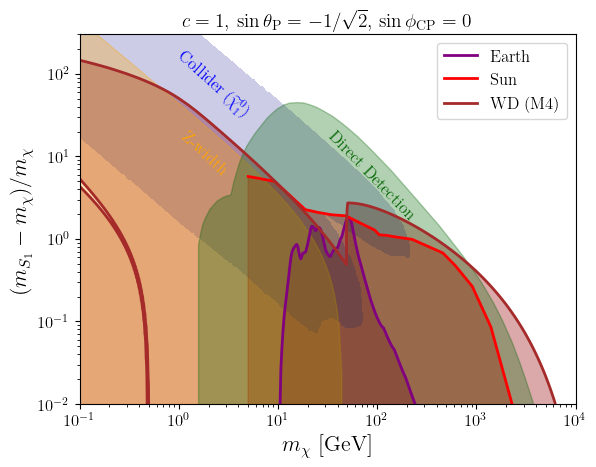}
    \caption{Excluded regions of a simplified model consisting of a Dirac dark matter particle of mass $m_\chi$ that couples to the tau lepton and a scalar mediator of mass $m_{S_1}$, for $c=1$, $\sin\theta_{\rm P}=1/\sqrt{2}$ and $\sin\phi_{\rm CP}=1$ (left panel) or $\sin\theta_{\rm P}=-1/\sqrt{2}$ and $\sin\phi_{\rm CP}=0$ (right panel). }
    \label{fig:exclusion_Dirac_Tau}
\end{figure}

In \cref{fig:exclusion_Dirac_Tau} we present the exclusion limits on the relative mass-splitting $(m_{S_1}-\mDm)/\mDm$ as a function of $\mDm$ for Earth, Sun and WD together with constraints from the direct detection experiment LZ (taken from Ref.~\cite{Ibarra:2024mpq}), SUSY stau searches at colliders~\cite{LEPSUSYWG_04011,ATLAS:2024fub} and model-independent constraints from the $Z$-width measurement~\cite{ALEPH:2005ab}. In the left panel, we assumed $c=1$, $\sin\theta_\text{P} = 1/\sqrt2$ and $\sin\phi_\text{CP}=1$, which enhances the magnetic moment, electric dipole moment and charge radius.   The ``break'' for WD at $\mDm=50\gev$ originates from the choice of setting $\mu_\chi = d_\chi = 0$ for $\mDm < 50\gev$, avoiding the region in which collective effects are expected to be relevant for DM capture in WDs~\cite{DeRocco:2022rze}. Therefore, for masses below this threshold, we only consider the constraints on the dimension-6 operators, thereby weakening the reach of WDs on the toy model parameter space. For this choice of parameters, the constraints from direct detection experiments are much stronger than those from capture in celestial bodies. However, this conclusion does not hold in the whole parameter space of the model.  In the right plot, we show the same figure but assuming  instead  $\sin\theta_\text{P} = -1/\sqrt2$ and $\sin\phi_\text{CP}=0$, which minimizes the scattering rate at $\mDm \gtrsim 7$ GeV at PICO-60, LZ, XENON1T and DS50. Here we find a region at large dark matter masses and small mass splittings where the heating constraints from heavy white dwarfs gives the strongest sensitivity in the parameter space. These plots illustrate the potential role of the capture in the celestial bodies, and subsequent annihilation into SM particles, for probing uncharted regions of the parameter space of dark matter models. 

\section{Majorana dark matter with a $t$-channel scalar mediator}
\label{sec:Majorana-toy-model}

The phenomenology of $\chi$ being Majorana changes drastically compared to the Dirac model discussed in the previous section, see \textit{e.g.} Ref.~\cite{Garny:2015wea}. The Majorana condition restricts the number of possible diagonal electromagnetic form factors to be one, such that $\chi$ can couple to the photon exclusively via the anapole moment $\ana_\chi$, whose value is twice as large as in the Dirac case~\cite{Schechter:1981hw,Pal:1981rm,Nieves:1981zt}.

\begin{figure}
    \centering
    \includegraphics[width=0.49\linewidth]{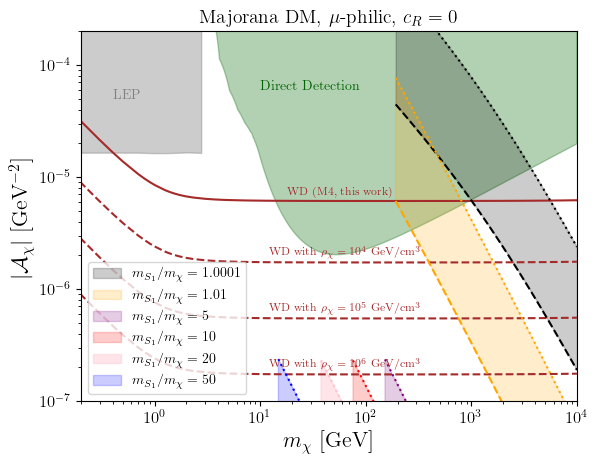}
    \includegraphics[width=0.49\linewidth]{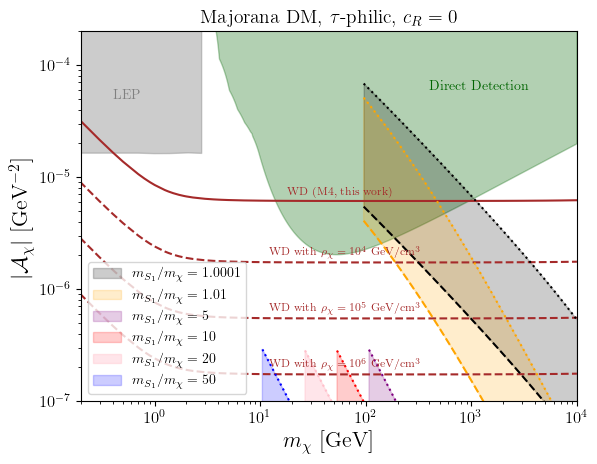}
    \includegraphics[width=0.49\linewidth]{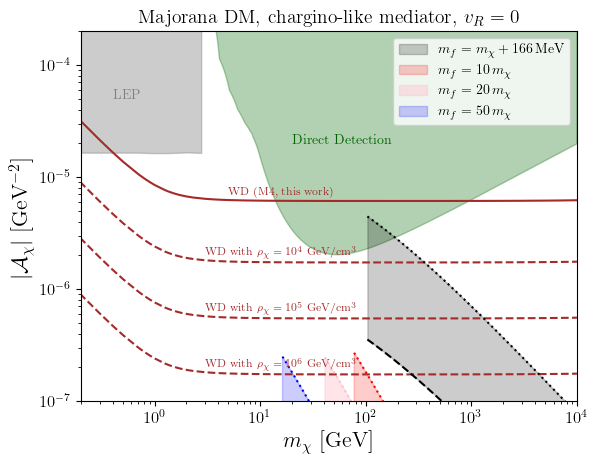}
    \caption{Scalar (top row) and vector (bottom) theory predictions for the anapole moments of a Majorana DM candidate. The upper- and lower edges of the theoretical predictions correspond to the choice of $\{c_L, v_L\}=\sqrt{4\pi}$ and $\{c_L, v_L\} =1$, respectively. We also show our constraints derived from the M4 WD (brown, continuous), and estimates for future observations of WDs in overdense DM regions (brown, dashed).}
    \label{fig:exclusion_Majorana}
\end{figure}

We show in \cref{fig:exclusion_Majorana} our WD constraints and model predictions for a $\mu$-philic (left) and tau-philic (right) toy model, in which the anapole moment is generated via scalar particles $S_1$ coupling to $\mu_L$ or $\tau_L$ as per \cref{eq:anapole_moment}. As a lower cut-off for the scalar mass, we chose the model-independent constraint, $m_{S_1}\gtrsim 45\gev$, inferred from the $Z\to\text{inv.}$ width~\cite{ALEPH:2005ab}. We further implemented collider constraints from LEP~\cite{LEPSUSYWG_04011} and ATLAS SUSY searches on smuons~\cite{ATLAS:2019lng,ATLAS:2019lff,ATLAS:2022hbt,ATLAS:2025evx} and staus~\cite{ATLAS:2024fub}, effectively excluding light DM candidates from the spectrum, while allowing the co-annihilation region of small mass-splittings due to the soft production at colliders. The different colors correspond to different values for the parameter $\eta = m_{S_1}/\mDm$, see the legend, and the upper- and lower edges assume $v_L=\sqrt{4\pi}$ and $v_L=1$, respectively. We find that large values for $\eta$ allow light DM to have sizeable anapole moments. In principle, the measurement of the $(g-2)_\mu$ comprises an additional constraint on the parameter space of the smuon-like toy model~\cite{Bringmann:2012vr}. 

Meanwhile, very small mass splittings can enhance the anapole moment of heavy DM $\mDm\gtrsim1\tev$, in particular if DM couples to a light lepton. For heavy DM candidates, the aforementioned constraints on the mediator are avoided, allowing current WD observations to constrain highly mass-degenerate models via their large predictions for the DM anapole moment. 

Future luminosity measurements of old WDs could alter this picture. We show in \cref{fig:exclusion_Majorana} an estimate of the exclusion constraints of a WD found in an region of DM overdensity with $\rho_\chi \in [10^4, 10^5, 10^6]\gev\,\text{cm}^{-3}$ assuming the same luminosity, WD mass and DM velocity distribution as for our analysis of the M4 WD. As the constraints on the electromagnetic moments scale as $\rho_\chi^{-1/2}$, relatively large DM overdensities are required to achieve a sizeable boost in the detection sensitivity. Obtaining a similar improvement with direct detection experiments would require large exposures, and a delicate analysis as the experiments are starting to probe the neutrino fog~\cite{OHare:2021utq}.

In principle, the DM anapole moment can also be generated by an interaction of the type
\begin{equation}
    \Lag \supset \bar\chi \gamma^\mu \left[v_L P_L + v_R P_R \right] f V_\mu + \hc,
\end{equation}
where $f$ is a charged Dirac fermion, $V$ is a charged vector boson, and $v_{L(R)}$ are the complex couplings parameterizing the interaction strenght between $\chi$ and $f_{L(R)}$. Note that in specific models, such as the MSSM, Goldstone bosons may also contribute to the anapole moment at the one-loop level and thus should be added. For our estimate here, we do not include such terms. The anapole moment then can be approximated as~\cite{Ibarra:2022nzm}
\begin{equation}
    \ana_\chi = \frac{e}{48\pi^2\mDm^2} \left[|v_L|^2 - |v_R|^2\right]\mathcal{F}_V\left(\frac{m_f}{\mDm},\frac{m_V}{\mDm}\right),
\end{equation}
where $\mathcal{F}_V$ is a loop function, see \cref{app:loop_functions}. For a parametric estimate, we assume $V=W$, \textit{i.e.} $\chi$ is neutralino-like, and $f$ is chargino-like. In the bottom panel of \cref{fig:exclusion_Majorana} we show the prediction for the anapole moment assuming the canonical electroweak mass-splitting of $m_f - \mDm = 166\mev$~\cite{Cheng:1998hc,Feng:1999fu,Gherghetta:1999sw} in black, as well as a large mass-splitting of $m_f/\mDm =$ $10$, $20$, $50$ in green, pink, and blue respectively. We only show points satisfying the constraints of LEP~\cite{LEPSUSYWG_01031}, restricting the chargino-like mass to $m_f > 103.5\gev$, as well as the constraints obtained from chargino-pair production at ATLAS~\cite{ATLAS:2024qxh}. For the latter, we assume no next-to-lightest neutralino in the spectrum. If such an additional dark sector particle is present, the exclusion limits could cover smaller mass-splittings, resulting in weaker predictions for the anapole moment in the bottom panel of \cref{fig:exclusion_Majorana} of the DM candidate. We find that the M4 WD constraints can probe new parameter space if there are additional enhancement factors, such as from $N$ species of chargino-like fermions in the loop. For example, for small mass-splittings and a chiral coupling $(v_L,v_R)=(v_L,0)$, the anapole moment can be approximated as
\begin{equation}
    \ana_\chi \sim \left(\frac{N}{10}\right) \left(\frac{|v_L|}{\sqrt{4\pi}}\right)^2\left(\frac{1\tev}{\mDm}\right)^2 \times 10^{-5}\gev^{-2},
\end{equation}
a prediction that would be testable with current M4 WD data for $N=10$, $|v_L|=\sqrt{4\pi}$ and $\mDm =1\tev$.

\section{Conclusions}\label{sec:Conclusion} 
In this work, we have investigated celestial bodies as probes of the electromagnetic interactions of spin-$1/2$ dark matter. We have calculated the capture of dark matter particles in the Sun, Earth, Jupiter, and massive white dwarfs through their millicharge, magnetic and electric dipole moments, charge radius, and anapole moment. The subsequent annihilation of the captured particles can generate observable neutrino fluxes from the Sun and Earth, or contribute to the heating of Jupiter and white dwarfs. Using current neutrino and luminosity data, we have derived constraints on the individual electromagnetic moments and compared them with existing bounds from direct dark matter searches, laboratory experiments, and astrophysical observations. We find that celestial bodies provide leading constraints in several regions of parameter space. In particular, the Earth is a powerful probe of millicharged dark matter, while massive white dwarfs provide strong constraints on the charge radius and anapole moment. Importantly, celestial bodies remain sensitive to large electromagnetic interactions for which the dark matter flux reaching underground direct-detection experiments is strongly attenuated by the terrestrial overburden.

We have also illustrated the implications of these constraints in simplified models in which the electromagnetic moments are radiatively generated. For Dirac dark matter coupled to charged scalar mediators, we find that for small CP-violation white dwarfs can probe new parameter space, setting leading constraints on heavy dark matter with small mass splittings to the scalar mediator. Elsewhere, white dwarfs are complementary to collider searches at low dark matter masses, while the remaining celestial bodies do not improve on existing constraints. The situation is particularly interesting for Majorana dark matter, for which the only allowed diagonal electromagnetic form factor is the anapole moment. Current observations of massive white dwarfs can probe new regions of parameter space for dark matter masses below approximately $10~{\rm GeV}$ and above the TeV scale, including highly mass-degenerate regions that are challenging for collider and direct-detection searches. Future observations of sufficiently massive and cold white dwarfs in regions of enhanced dark matter density could substantially extend this sensitivity. These results demonstrate the complementarity between celestial-body capture and terrestrial searches in probing the electromagnetic properties of dark matter.

\acknowledgments
The work of MR is supported by the KIAS Individual Grant PG108401 and by the Center for Advanced Computation (CAC) at Korea Institute for Advanced Study. 
The work of AI is supported by the Collaborative Research Center SFB1258 and by the Deutsche Forschungsgemeinschaft (DFG, German Research Foundation) under Germany's Excellence Strategy - EXC-2094 - 390783311. MR and GT are thankful to the organizers of the 2026 MIAPbP
workshop “Fill the Gap” for their hospitality and acknowledge the support of the Munich Institute for Astro-, Particle and BioPhysics (MIAPbP) which is funded by the Deutsche Forschungsgemeinschaft (DFG, German Research Foundation) under Germany's Excellence Strategy – EXC-2094 – 390783311.

\appendix

\section{Loop-induced electromagnetic moments}\label{app:loop_functions}
In this Appendix, we briefly summarize the analytical formulas of the electromagnetic moments of a neutral spin-1/2 fermion that are generated by a $t$-channel mediator at the one-loop level. The results are taken from Refs.~\cite{Ibarra:2022nzm,Ibarra:2024mpq}.

For a $t$-channel scalar mediator (with mass $m_S$), a Lagrangian of the type
\begin{equation}
    \Lag = \bar\chi [c_L P_L + c_R P_R] S^* f + \hc
\end{equation}
generates all dimension-full electromagnetic moments. The fermion with charge $Q_f$ has a mass denoted by $m_f$, and $c_{L,R}$ are complex coupling constants. The magnetic dipole moment is given by
\begin{equation}
\mu_\chi = \frac{-eQ_f}{32\pi^2m_\chi}\biggr\{(|c_L|^2+|c_R|^2)\,\mathcal{F}_1 \left(\frac{m_f}{m_\chi},\frac{m_{S}}{m_\chi}\right)+2 \Re{c_L c_R} \,\mathcal{F}_2\left(\frac{m_f}{m_\chi},\frac{m_{S}}{m_\chi}\right)\biggr\},
\end{equation}
with loop functions
\begin{align}
\mathcal{F}_1(\mu,\eta) =&-1+\frac{1}{2}(\mu^2-\eta^2)\log(\frac{\mu^2}{\eta^2})\nn\\
& -\frac{(\eta^2-1)(\eta^2-2\mu^2)-\mu^2(3-\mu^2)}{\sqrt{\Delta}}\arctanh{\frac{\sqrt{\Delta}}{\eta^2+\mu^2-1}}
\end{align}
and
\begin{align}
\mathcal{F}_2(\mu, \eta) = \mu \left[\frac{1}{2}\log(\frac{\mu^2}{\eta^2})+\frac{\eta^2-\mu^2+1}{\sqrt{\Delta}}\arctanh{\frac{\sqrt{\Delta}}{\eta^2+\mu^2-1}}\right],
\end{align}
with $\Delta = (\mu^2-\eta^2 + 1)^2 - 4\mu^2$.
The electric dipole reads
\begin{equation}
d_\chi =\frac{eQ_f}{16\pi^2m_\chi} \Im{c_L c_R} \mathcal{F}_2\left(\frac{m_f}{m_\chi}, \frac{m_S}{m_\chi}\right),
\end{equation}
and the anapole moment is given by
\begin{equation}\label{Anapole:eq:Fscalar}
\mathcal{A}_\chi =- \frac{e Q_f}{192\pi^2\mDm^2}\left[|c_L|^2-|c_R|^2\right]\mathcal{F}_3\Big(\frac{m_f}{m_\chi},\frac{m_S}{m_\chi}\Big)\;
\end{equation}
with
\begin{equation}
\mathcal{F}_3(\mu,\eta)=\frac{3}{2}\log(\frac{\mu^2}{\eta^2})+\frac{3\eta^2-3\mu^2+1}{\sqrt\Delta} \arctanh{\frac{\sqrt{\Delta}}{\eta^2+\mu^2-1}}.
\end{equation}
Finally, the charge radius is given by
\begin{equation}
b_\chi = \frac{-e Q_f }{384 \pi^2 m_\chi^2}\left[(|c_L|^2+|c_R|^2) \mathcal{F}_4\left(\frac{m_f}{m_\chi},\frac{m_{S}}{m_\chi}\right) + 2 \Re{c_L c_R}  \mathcal{F}_5\left(\frac{m_f}{m_\chi},\frac{m_{S}}{m_\chi}\right) \right],
\end{equation}
where we defined
\begin{align}
\mathcal{F}_4(\mu,\eta) &= \frac{2\left( 8\Delta^2 + \Delta(9\eta^2+7\mu^2-5)-4\mu^2(3\eta^2+\mu^2-1)\right)}{\Delta^{3/2}} \arctanh{\frac{\sqrt{\Delta}}{\eta^2+\mu^2-1}} \nn \\
&+\frac{4(4\Delta+\eta^2+3\mu^2-1)}{\Delta} + (8\mu^2-8\eta^2-1)\log(\frac{\eta^2}{\mu^2}),
\end{align}
and
\begin{align}
\mathcal{F}_5(\mu,\eta)  = 8 \mu \biggr[&\frac{\Delta+\eta^2 (-\Delta+2 \mu^2+1)+\mu^2 (\Delta-2 \mu^2+3)-1 }{\Delta^{3/2}} \arctanh{\frac{\sqrt{\Delta}}{\eta^2+\mu^2-1}}\nn \\
&+ \frac{\mu^2-\eta^2}{\Delta} + \frac{1}{2} \log(\frac{\eta^2}{\mu^2}) \biggr].
\end{align}

Further, a vector mediator $V$ can also induce electromagnetic moments at the one loop level via a Lagrangian
\begin{equation}
    \Lag = \bar\chi\gamma^\mu \left[v_L P_L + v_R P_R\right] f V_\mu^+ + \hc
\end{equation}
with complex couplings $v_{L,R}$ and charged fermion $f$ (with mass $m_f$).
In principle, also the degrees of freedom associated with the longitudinal vector degree of freedom can contribute to the electromagnetic moments of $\chi$. This model-dependent contribution can be calculated using the results for the scalar mediator with $m_S=m_V$. The contribution to the anapole moment reads
\begin{equation}
    \ana_\chi = \kappa \frac{-eQ_f}{96\pi^2\mDm^2} \left[|v_L|^2 - |v_R|^2\right]\mathcal{F}_V\left(\frac{m_f}{\mDm},\frac{m_V}{\mDm}\right),
\end{equation}
with $\kappa=2$ ($\kappa = 1$) if $\chi$ is a Majorana (Dirac) fermion. The loop function reads
\begin{equation}
    \mathcal{F}_V(\mu,\eta) = \frac{3}{2} \log\left(\frac{\mu^2}{\eta^2}\right) + \frac{3\eta^2 - 3\mu^2 - 7}{\sqrt{\Delta}} \arctanh\left(\frac{\sqrt{\Delta}}{\eta^2 + \mu^2-1}\right).
\end{equation}


\providecommand{\href}[2]{#2}\begingroup\raggedright\endgroup

\end{document}